\documentclass[reprint, 10pt, floatfix,aps,twocolumn,a4paper,superscriptaddress,nofootinbib,notitlepage,pra,letterpaper]{revtex4-1}
\usepackage{inputenc}
\usepackage[T1]{fontenc} 
\usepackage{paralist}
\usepackage{hyperref}
\usepackage{amsopn,amsthm,dsfont,amssymb}
\usepackage{mathrsfs}
\usepackage{cancel} 
\usepackage{color} 

\usepackage{soul} 
\usepackage[normalem]{ulem}
\usepackage{cancel} 
\usepackage{hyperref} 
\usepackage{soul,xcolor}
\usepackage{lipsum}
\usepackage[english]{babel}

\hypersetup{
    colorlinks=true,
    linkcolor=blue,
    filecolor=magenta,      
    urlcolor=cyan,
}

\AtBeginDocument{
    \newwrite\bibnotes
    \def\bibnotesext{Notes.bib}
    \immediate\openout\bibnotes=\jobname\bibnotesext
    \immediate\write\bibnotes{@CONTROL{REVTEX41Control}}
    \immediate\write\bibnotes{@CONTROL{%
    apsrev41Control,author="08",editor="1",pages="1",title="0",year="1"}}
     \if@filesw
     \immediate\write\@auxout{\string\citation{apsrev41Control}}
    \fi
}

\newcommand{\ket}[1]{|#1\rangle}

\usepackage{mathtools}

\usepackage{adjustbox}
\usepackage[caption=false]{subfig}

\begin{document}
\title{The Gravito-Phononic Effect: A Quantum Signature of Linearised Gravity}

\author{Germain Tobar}
\affiliation{Department of Physics, Stockholm University, SE-106 91 Stockholm, Sweden}

\author{Oscar Berg}
\affiliation{Department of Physics, Stockholm University, SE-106 91 Stockholm, Sweden}

\begin{abstract}
    The photo-electric effect was a historic milestone in the development of quantum theory, revealing the first evidence of discrete energy of the electromagnetic field through hallmark signatures such as the threshold frequency, intensity-independent energy transfer, and the near instantaneous ejection of photo-electrons. Here, we discuss the photo-electric effect through the lens of semi-classical, quantum, and neo-classical models. We provide a pedagogical outline for how two coupled harmonic oscillators, under a beam-splitter interaction, can exhibit hallmark signatures analogous to the photo-electric effect, including resonance conditions and quantised energy absorption. We discuss the implications of this model for recently proposed graviton detection protocols. This further clarifies that a gravitational version of the photo-electric effect, modelled as discrete energy transfer between harmonic oscillators can provide the first evidence of the graviton.
\end{abstract}

\def\thefootnote{}\footnotetext{germain.tobar@fysik.su.se}
\maketitle
\section{Introduction}
The detection of gravitational radiation has opened a new era in astronomy, with signals observed from binary black hole and neutron star mergers providing profound insights into the cosmos \cite{abbott2016observation,AbbottB.P.2016SotA,abbott2017gw170817,abbott2019narrow,abbott2023open}. Concurrently, advancements in controlling quantum systems at macroscopic scales, such as the preparation of non-classical states of mechanical oscillators, have pushed the boundaries of quantum sensing technologies \cite{WollmanE.E.2015Qsom,ChuYiwen2017Qaws,SatzingerKJ2018Qcos}. These developments have spurred interest in exploring quantum aspects of gravity, where novel proposals aim to detect quantum features of gravitational source masses  \cite{bose2024massivequantumsystemsinterfaces,bose2017spin,marletto2017gravitationally,2022fqce.book...85A}, as-well as on-shell gravitational radiation \cite{BlencoweMP2013Efta,PhysRevD.98.124006, parikh2020noise,kanno2021noise, Guerreiro2022quantumsignaturesin,manikandan2024detectingacoherenceradiationfields}.

Recent theoretical proposals show that contrary to the analysis of previous works \cite{BoughnStephen2006Aogd,DYSONFREEMAN2013IAGD}, it is be possible to detect single gravitons-the hypothetical particles of the gravitational field with macroscopic acoustic oscillators \cite{tobar2023detecting, tobar2024detectingkhzgravitonsneutron}, or essentially any gravitational wave detector with read-out in particle number basis \cite{carney2023}. More precisely, by combining phononic gravitational wave detectors \cite{Weberoriginal,astone_resonant_1997} with modern phonon counting techniques \cite{HongSungkun2017HBaT, VelezSantiagoTarrago2019PaDo, MirhosseiniMohammad2020Sqto,vonLüpkeUwe2022Pmit}, it is possible to count single gravitons. In essence, all it takes is a gravitational wave detector, with read-out in particle number basis - the ratio of number of discrete energy transitions in the detector to the number of incident quanta in the field, then defines a quantum efficiency for the graviton detector \cite{carney2023}. The graviton detection schemes in Refs.~\cite{tobar2023detecting, tobar2024detectingkhzgravitonsneutron,carney2023}, follow analogously to the historic photo-electric, and mimic electromagnetic photo-detection processes \cite{tobar2023detecting,carney2023}, which provided early evidence for the quantisation of electromagnetic radiation, although not a smoking-gun proof \cite{carney2023}. For further details of the historical context of the photo-electric effect, we refer to Ref.~\cite{shenderov2024stimulatedabsorptionsinglegravitons}. 

Here we provide a pedagogical comparison of phononic and atomic systems as detectors of energy of radiation involving energy transfer between coupled harmonic oscillators. By examining semi-classical, quantum, and neo-classical models of the photo-electric effect, we outline that a system of two coupled oscillators (which as we outline forms a generic model for all the distinct graviton detection schemes proposed in Refs.~\cite{tobar2023detecting, tobar2024detectingkhzgravitonsneutron,carney2023}), can replicate the core signatures of discrete energy absorption observed in photo-electric transitions. This includes the presence of a threshold frequency, intensity-independent energy transfer, and rapid transition times.

We discuss how this framework applies to the context of linearized quantum gravity, where the graviton-phonon interaction serves as an analog to photon-electron interactions in quantum electrodynamics, corrobarating the connections to the photo-electric effect for gravito-phononic transitions made in Ref.~\cite{tobar2023detecting}. More generally, the coupled oscillator model is a generic model for the energy exchange for all recently proposed graviton detection methods \cite{tobar2023detecting, tobar2024detectingkhzgravitonsneutron,carney2023}.

\section{The photo-electric effect}\label{Xsec2-2} \label{PEeffect1}
\begin{figure}
\begin{center}
    \includegraphics[width=0.92\linewidth]{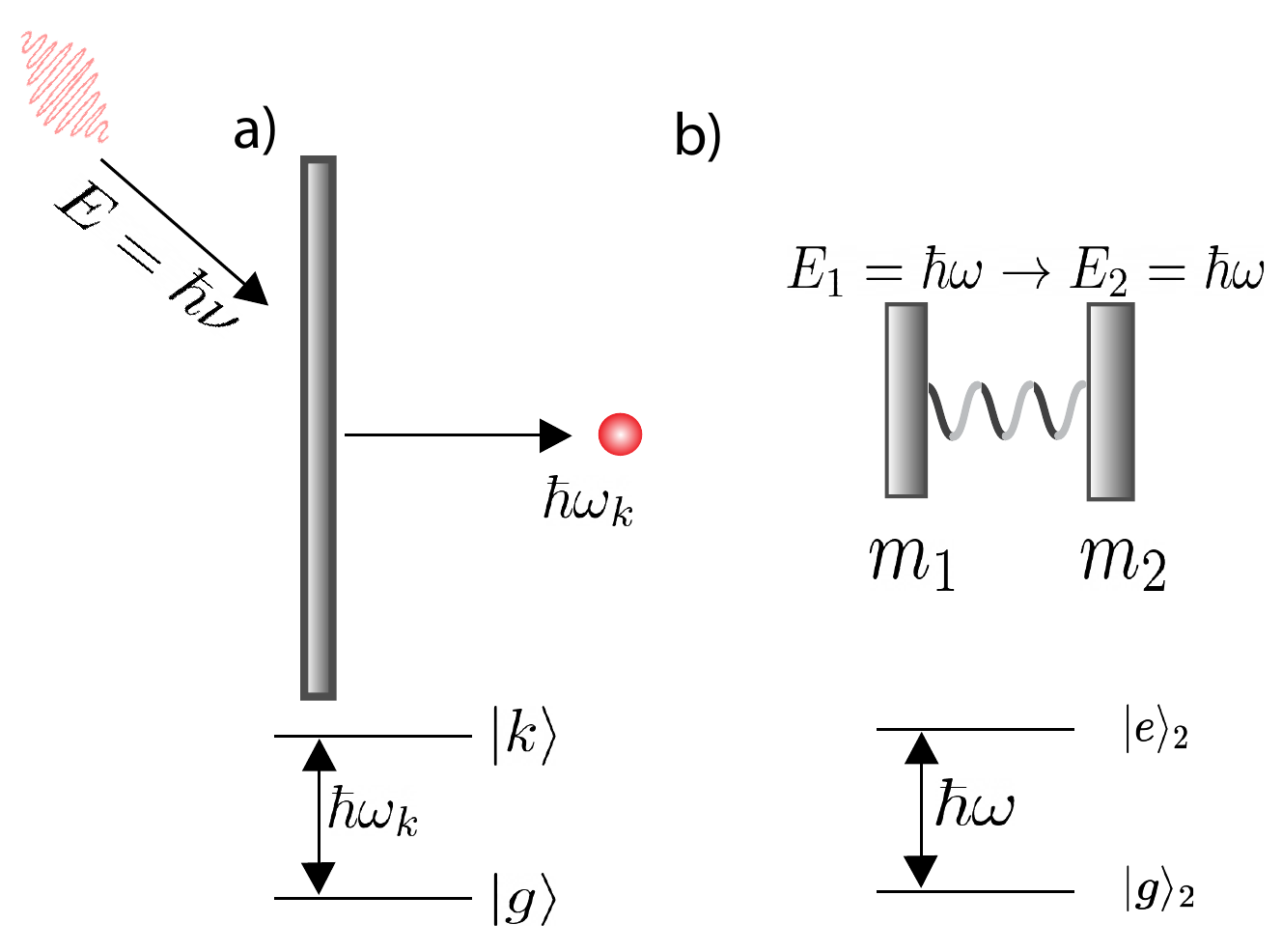}
  \caption{The requirements for $\hbar \nu$ energy exchange in photon to electron conversion for photo-electric transitions in a) is also required for $\hbar \omega$ energy exchange between two coupled harmonic oscillators in b). In a gravitational context, the gravitational field plays the role of one bosonic mode, while either the resonant mass or interferometric detector plays the role of the second bosonic mode.}
  \label{exchange plot}    
\end{center} 
\end{figure}

Historically, three key signatures of the photoelectric effect were interpreted as evidence for the quantisation of the electromagnetic field \cite{lamb1968photoelectric}. These include a) the existence of a threshold frequency, b) the independence of the energy of the ejected electron from the intensity of the electromagnetic field, and c) the near instantaneous ejection of photo-elecrons. We now provide a pedagogical outline of how each of these signatures can be explained in both hybrid semi-classical models and full quantum electrodynamics (Fig.{\ref{exchange plot}}).

\subsection{Semi-classical theory and quantum electrodynamics}\label{Xsec3-2.1}\label{scem} A common toy model for energy exchange between an atomic detector is a harmonic oscillator interacting with a qubit:
\begin{equation}\label{Hxxpp}
    H=\frac{1}{2} x^2+\frac{1}{2} p^2+ \hbar\omega\sigma_z +\lambda x\sigma_x ,
\end{equation}
where here $x$ and $p$ remain classical operators satisfying $\{x, p\}=1$, while the pauli matrices satisfy their standard commutation relations. Using the model in Eq.~{\eqref{Hxxpp}}, in the regime of weak couplings, perturbation theory demonstrates that the probability for the qubit to transition to the excited state is \citet{lamb1968photoelectric}:
\begin{equation}
   P_e =  \frac{\lambda^2 \sin ^2((\omega -\nu) t / 2)}{(\hbar(\omega -\nu))^2},
\label{Xeqn2-2}
\end{equation}
and here we see the archetypal threshold frequency, if the frequency of the incident radiation is lower than the threshold frequency $\omega$, then the probability for the electron to be excited is suppressed. The next signature, is that the energy of the transition is independent of the amplitude of the electromagnetic field. Finally, at arbitrarily small times, there is a non-zero probability for the electron to be excited. Therefore, the semi-classical interaction of a classical harmonic oscillator, with a qubit can replicate all core signatures of the photo-electric effect as a resonance, as has been discussed in Refs.~\cite{tobar2023detecting,PhysRevD.109.044009,shenderov2024stimulatedabsorptionsinglegravitons,carney2024commentsgravitondetection}. Extension to excitations into a continuum of excited state energy levels, can be made as in Ref.~\cite{lamb1968photoelectric}, by considering the excitation as being into a state in a continuum of excited states $\ket{k}$, with energy $\hbar\nu_k$.

If Eq.~{\eqref{Hxxpp}} is taken at face value, the total energy before and after completing the $\hbar \omega$ energy transition is not conserved. In order to see this, we can re-write Eq.~{\eqref{Hxxpp}}  as the standard interaction between an atomic detector and the electromagnetic field in the Coloumb gauge:
\begin{equation}\label{Hfff}
\begin{split}
      \hat{H}_{\mathrm{SC}} &=  \frac{1}{2} \int d^3 x\left(\dot{\mathbf{A}}_T(x)^2+(\nabla \times \mathbf{A}_T(x))^2\right)\hat{I} + \hbar\frac{\omega}{2} \sigma_z \\&\ \ + \hbar g_{\mathrm{SC}}(\sigma_{-}+ \sigma_{+}), 
\end{split}
\end{equation}
where $\mathbf{A}_T$ is now a c-number classical field for the transverse component of the electromagnetic vector potential rather than an operator, and $g_{\mathrm{SC}}$ is the semi-classical interaction strength between the classical field and the detector. Here, Eq.~{\eqref{Hfff}}, is the semi-classical theory of classical electromagnetic radiation interacting with a quantised detector. Although we have explicitly added the free-energy of the electromagnetic field $H_F~=~\frac{1}{2} \int d^3 x\left(\dot{\mathbf{A}}_T(x)^2+(\nabla \times \mathbf{A}_T(x))^2\right)\hat{I}$, such that the expectation value of \eqref{Hfff}, gives the joint energy of the classical electromagnetic field and quantised detector. Historically, despite this classical explanation, the photo-electric effect was still treated as evidence of photons due to the non-equivalence of $H_F$ and the energy absorbed by the detector $\hbar \omega$ \cite{Clauser74,MandelLeonard1995Ocaq}. In other words, if the free-energy of the electromagnetic field $H_F$, remains unchanged after the detector undertakes a transition in energy by $\hbar \omega$, from $\ket{g} \rightarrow \ket{e}$, then energy conservation is violated. In the full quantum field theory, the Hamiltonian is
\begin{equation}\label{combinedHem}
  \hat{H} = \hbar\nu a^{\dagger} a  +   \hbar\frac{\omega}{2} \sigma_z + g_{q, \nu}(a\sigma_- + a^{\dagger}\sigma_+), 
\end{equation}
where $g_{q, \nu}$ is the coupling strength to the electromagnetic field vacuum, at frequency $\nu$. In the QFT, energy conservation is restored as $H_F$ becomes $\hbar\nu a^{\dagger} a$, which can now account for photo-electric transitions through the addition or subtraction of single quanta of energy $\hbar \nu$.

\subsection{Energy conservation in photo-electric transitions}\label{Xsec4-2.2}
We now flesh out in more detail the energy conservation argument originally discussed qualitatively in Ref.~\cite{tobar2023detecting}. The semi-classical or quantum Rabi model for the exchange of energy between a field and a detector, gives the following generic expression for the transition probability of an atom driven by an electromagnetic field:
\begin{equation}
    P_e = \frac{g^2}{g^2 + \delta^2}\sin^2\left(\frac{(\delta^2 + g^2)t}{2}\right),
\label{Xeqn5-5}
\end{equation}
where $\delta = \omega_0 -  \nu$ is the detuning between the atom transition frequency and the frequency of the incident electromagnetic field, $g = \sqrt{n}g_{q,\nu}$ is the coupling strength between the field and the atom enhanced from the vacuum coupling $g_{q,\nu}$ by the square root of the number of photons in the incident field (for $n \gg 1$, $g_{SC} = g$). This expression in the quantum Rabi model gives the transition probability between energy eigenstates of the free Hamiltonian $\ket{n}\ket{g}$ and $\ket{n-1}\ket{e}$. Here, it is clear that there can be a substantial probability for non-energy conserving transitions (between free Hamiltonian eigenstates), particularly for large coupling strengths $g_\mathrm{cl}$, i.e. transitioning between these free energy eigenstates corresponds to a violation of energy conservation by $\mathrm{E}_\mathrm{diff} = \hbar \delta$. However, the spread of the energy of the joint system in either of the eigenstates discussed above is given by 
\begin{equation}
  \Delta E  = \mathcal{O}\left(\sqrt{\mathrm{Var}(\hat{H})}\right) = \mathcal{O}(g_\mathrm{cl}). 
\label{Xeqn6-6}
\end{equation}
Now, in the regime in which the energy non-conservation becomes substantially larger than the energy variance (given by the coupling strength), such that $\delta \gg g_\mathrm{cl}$, then the expression for the probability of the transition discussed above reduces to:
\begin{equation}
    P_e = \frac{g^2}{\delta^2}\sin^2\left(\frac{\delta^2 t}{2}\right).
\label{Xeqn7-7}
\end{equation}
This is the well-known Fermi-Golden rule limit, which for long times approaches the Dirac-delta function (in the limit of infinite times this probability becomes zero for $\delta \neq 0$). We see here that for long times and weak coupling strengths the probability of non-energy conserving transitions are statistically suppressed $P_e \rightarrow 0$. In the same regime, $E_\mathrm{diff} = \hbar  \delta$. It is in exactly this sense that the QFT preserves energy conservation, while the CFT does not - while the CFT predicts the same transition probability, the CFT maintains that the difference in energy between the free-energy eigenstates approaches $E_\mathrm{diff} \rightarrow \hbar \nu$, violating energy conservation by $\hbar \nu$. However, the classical field theory does satisfy an ensemble average energy conservation law, in that the time-independence of the Hamiltonian directly implies that the ensemble average of the Hamiltonian is a conserved quantity. However, energy conservation at the level of single transitions is violated, as shown here.

\subsection{Neo-classical theory}\label{Xsec5-2.3}
While the semi-classical theory of radiation described in Section~\ref{scem}, can account for all core signatures of the photo-electric effect, it doesn't account for the back-reaction of the field on the quantum matter. A simple model which accounts for this can be built as a non-linear model \cite{PhysRevD.37.3522} with the same Hamiltonian in Eq.~{\eqref{Hxxpp}}, but with time evolution of the classical variables generated through \cite{terno2024classicalquantumhybridmodels}
\begin{equation}
\begin{split}
\dot{x}&=\{x,\langle\hat{H}\rangle\}=\partial_p\langle\hat{H}\rangle \\
\dot{p}&=\{p,\langle\hat{H}\rangle\} = -\partial_x\langle\hat{H}\rangle,
\end{split}
\label{Xeqn8-8}
\end{equation}
while generating time evolution of the wavefunction of the detector through 
\begin{equation}
i \hbar \frac{d \psi}{d t}=i \hbar\left(\partial_t \psi+\left\langle\boldsymbol{\xi}_{\hat{H}}\right\rangle \cdot \nabla_z \psi\right)=\hat{H} \psi,
\label{Xeqn9-9}
\end{equation}
where $\langle\hat{H}\rangle=\left\langle\psi\right| \hat{H}\left|\psi\right\rangle$ and $\left\langle\boldsymbol{\xi}_{\hat{H}}\right\rangle=\left(\partial_p\langle\hat{H}\rangle,-\partial_x\langle\hat{H}\rangle\right)$.  

In this way, the equations of motion for the classical variables, the equation of motion for the energy of the classical operator $E_F = \frac{p^2}{2} + \frac{x^2}{2}$ is:
\begin{equation}
    \begin{aligned}
\dot{E}_F & = - p\lambda \langle \sigma_x\rangle.
\end{aligned}
\label{Xeqn10-10}
\end{equation}
Importantly, this elevates the total energy of the classical field to a time-dependent quantity, rather than the usual assumption that the energy of the classical oscillator remains unchanged in the semi-classical theory. While this is not to the most general or least-pathological classical-quantum model for energy transfer between a classical oscillator and a qubit (we refer to Refs.~\cite{PhysRevX.13.041040,OppenheimJonathan2023Gidv,layton2023healthiersemiclassicaldynamics} for the development of consistent quantum-classical hybrid models), it motivates that hybrid models provide a pathway to minimising violations of energy conservation, at the expense of a non-linearity.

\section{Energy transfer between harmonic oscillators and the photo-electric effect}\label{Xsec6-3} 
We have examined how the historic photo-electric effect, modelled as a qubit coupled to the electromagnetic field, can be explained with quantum electrodynamics, or the semi-classical theory of radiation. In this section, we demonstrate how energy transfer between two coupled harmonic oscillators (rather than a harmonic oscillator and a qubit in the previous section), also exhibits the hallmark signatures of the photo-electric effect and discuss the differences to the qubit-bosonic mode coupling as describes the photo-electric transitions.

\subsection{Quantum Theory}\label{beamsplitterdynamics}
A simple model for energy transfer between two harmonic oscillators as bosonic modes in quantum theory, can be found through a beam-splitter interaction, in the rotating-wave-approximation:
\begin{equation}
    \hat{H} = \hbar \nu \hat{a}^\dag \hat{a} + \hbar \omega \hat{b}^\dag \hat{b} + \hbar g (\hat{a}\hat{b}^\dag + \hat{b}\hat{a}^\dag).
\label{Xeqn11-11}
\end{equation}
We can now evaluate the time-evolution of the probability for the detector mode (of frequency $\omega$). In Appendix.~\eqref{beamappendix}, we evaluate:
\begin{equation}
\begin{split}
    P(\ket{n = 1}) &= g^2|\alpha|^2\int_0^t d t^{\prime}\int_0^t d t^{\prime\prime} e^{-i(\omega-\nu) (t' -t'')} \\
    &= 4g^2|\alpha|^2\frac{\sin ^2\left(\frac{1}{2}\left(\nu-\omega\right) t\right)}{\left(\nu-\omega\right)^2},
\end{split}
\label{Xeqn12-12}
\end{equation}
which demonstrates the hallmark signatures of the photo-electric effect. The threshold frequency manifests as the resonance condition. While the energy of the produced phonon $\hbar\omega $ is independent of the intensity of the driving field. 

\subsection{Semi-classical Theory}\label{Xsec8-3.2}\label{sc111}
We can examine a semi-classical Hybrid model, that neglects the back-reaction for the quantum harmonic oscillator detector on the classical field as
\begin{equation}\label{h01}
        \hat{H}=\frac{1}{2} x^2+\frac{1}{2} p^2+\hbar \omega \hat{b}^\dag \hat{b}+\lambda x (\hat{b}^\dag + \hat{b}).
\end{equation}
Denoting $E_\mathrm{cl} = \frac{1}{2} x^2+\frac{1}{2} p^2$ as the energy of the classical oscillator, the total energy of the joint classical-quantum system can be estimated as the expectation value of 
\begin{equation}\label{Ecl}
        \hat{H}=E_\mathrm{cl}\hat{I} +\hbar \omega \hat{b}^\dag \hat{b}+\lambda x (\hat{b}^\dag + \hat{b}).
\end{equation}
Importantly, using Eq.~{\eqref{Ecl}} as the Hamiltonian for the dynamics of an initially ground state cooled quantum detector mode, the time-evolved state, is \cite{tobar2023detecting}:
\begin{equation}
    |\psi(t)\rangle=\left|\beta(t) e^{-i \omega_0 t}\right\rangle,
\label{Xeqn15-15}
\end{equation}
where the coherent state amplitude is $\beta(t)=-i \lambda \int_0^t d s \ddot{x}(s) e^{i \omega_0 s}$. If we consider that the classical variables follow their usual evolution under the Poisson brackets (such that there is no back-reaction of the classical field on the quantum detector), the equation of motion for the amplitude of the classical oscillator is $x(t) = x_0\mathrm{sin}(\nu t)$. In this case, the probability for the quantum detector to transition to the $n = 1$ Fock state is
\begin{equation}\label{Pn1}
    P(\ket{n = 1}) =  \lambda^2 x_0^2 \nu^4 \frac{t^2}{4} \operatorname{sinc}^2\left(\frac{\delta t}{2}\right),
\end{equation}
where $\delta = \nu - \omega$ is the detuning between the frequency of the classical oscillator and the transition frequency of the quantum harmonic oscillator detector. 

In analogy to the photo-electric signatures of Section~\ref{PEeffect1}, all core signatures of the photo-electric effect follow: the archetypal threshold frequency appears as the resonance condition, in which the frequency of oscillation of the classical oscillator is lower than the transition frequency of the detector, then the probability for the quantum detector to transition is suppressed. It is only when the frequency of the radiation, matches the frequency of the detector: 
\begin{equation}\label{einstpe}
    \hbar \nu = \hbar \omega,
\end{equation}
does the quantum detector have a non-negligible probability to absorb a single quanta of energy. This resonance condition, converted into the energy conservation relation of Eq.~{\eqref{einstpe}}, manifests as a phononic version of the Einstein photo-electric relation, which Millikan verified, using many harmonic oscillators of different frequencies and measuring discrete excitations with different gravitational wave frequencies will allow for a gravitational version of Millikan's experiment, this applies in the case of either resonant mass or any gravitational wave detector involving resonant energy exchange, including detectors utilising the electromagntetic field. This highlights how a simple model of a classical harmonic oscillator coupled to a quantum harmonic oscillator can replicate all core signatures of the photo-electric effect. However, as we highlight below, in full analogy to the case of photo-electric transitions, energy conservation is violated by $\hbar \omega$, for single quanta transitions if this model is maintained. 

In analogy with the photo-electric case, the probability of a non-energy conserving transition is statistically suppressed, as follows directly from Eq.~{\eqref{Pn1}}, for $\delta \gg \lambda$,  it also follows here that for long times and weak coupling strengths the probility of non-energy conserving transitions are statistically suppressed $P_e \rightarrow 0$. In the same regime, $E_\mathrm{diff} \rightarrow \hbar \delta$. Therefore, in full analogy to the photo-electric case, the QFT preserves energy conservation, while the CFT does not. The energy difference between the free-Hamiltonian eigenstates is $E_{\mathrm{diff}}=\hbar \delta$, with energy spread $\Delta E = \mathcal{O}(\lambda)$.

\subsection{Neo-classical theory}\label{Xsec9-3.3}
The theory of quantum matter back-reacting on a classical field can be accounted for in the harmonic oscillator modelling through using Eq.~{\eqref{h01}}, and considering the evolution:
\begin{equation}
    \begin{aligned}
\dot{x} & =\{x,\langle\hat{H}\rangle\} = p \\
\dot{p} & =\{p,\langle\hat{H}\rangle\} = - x - \lambda \langle \hat{b} + \hat{b}^\dag \rangle.
\end{aligned}    
\label{Xeqn18-18}
\end{equation}
In this Hybrid model, the evolution of the classical energy of the bosonic mode
\begin{equation}
\dot{E} = \{\frac{1}{2} x^2+\frac{1}{2} p^2,\langle\hat{H}\rangle\} = - \lambda p \left\langle\hat{b}+\hat{b}^{\dagger}\right\rangle,
\label{Xeqn19-19}
\end{equation}
is no longer constant, but depends on the expectation value of the amplitude quadrature of the quantised detector. This highlights how the back-reaction enables a partial restoration of energy conservation. However, this comes at the expense of a non-linearity.

\section{The gravito-phononic detector}\label{Xsec10-4}
We now outline explicitly how the previously outlined arguments apply to the case of a gravito-phononic detector. The semi-classical model for the interaction of gravitational radiation and matter can be modelled through the Hamiltonian:
\begin{equation}
\begin{split}
        & \hat{H} = \int_V\mathrm{d}^3x  \left(\frac{c^2}{32 \pi G}\right) \nu^2 h_0^2 \hat{I} + \hbar \omega \hat{b}^{\dagger} \hat{b}  + \frac{L}{\pi^2} \sqrt{\frac{M\nu^4 \hbar}{\omega_0}}\left(\hat{b}+\hat{b}^{\dagger}\right) h.     
\end{split}
\label{Xeqn20-20}
\end{equation}
The first term corresponds to the gauge-invariant energy density of the classical linearised gravitational field of a plane wave, averaged over several wavelengths, and integrated over some finite volume $V$. We can identify $\lambda = \frac{ML\nu^2}{\pi^2}$ as the coupling strength in the model from Section~\ref{sc111}, with $x_0 = \sqrt{\frac{\hbar}{M\omega_0}}$. 

Therefore it follows from the coupled harmonic oscillator model of Section~\ref{sc111}, that all core signatures of the photo-electric follow in this gravitational analogue. In analogy with the QED model, the model of the quantised gravitational field interacting with the detector is:
\begin{equation} 
    \hat{H}=\hbar \nu \hat{a}^{\dagger} \hat{a}+\hbar \omega \hat{b}^{\dagger} \hat{b}+\hbar g_{q,\nu }\left(\hat{a} \hat{b}^{\dagger}+\hat{b} \hat{a}^{\dagger}\right),  
\label{Xeqn21-21}
\end{equation}
where $g_{q,\nu } = \frac{1}{c} \sqrt{\frac{8 \pi G \hbar}{V \nu}}$ is the coupling strength to the vacuum of the $\nu$ frequency mode of the gravitational field. Therefore the model of Section~\ref{beamsplitterdynamics}, can be applied with the replacement $g \rightarrow g_{q,\nu }$, and in this way all core signatures of the photo-electric effect, including the requirement of linearised quantum gravity for a consistent model of the energy exchange are required. 

Furthermore, a similar beam-splitter interaction can be used in the case of gravitational radiation interacting with the electromagnetic detectors that use read-out in particle number basis due to resonant graviton to photon conversion, such as that proposed in Refs.~\cite{PhysRevLett.74.634,AlexanderDolgov,ADDAZI2024138574,ADDAZI2025101844}, would also in a similiar way correspond to a gravitational version of the photo-electric effect.

In summary, we re-iterate here that any experiment whether it is the photo-electric effect \cite{einstein1905erzeugung}, spontaneous emission \cite{MANDEL197627}, or even tests for co-incidence counts of single photons \cite{Clauser74} can be explained with some contrived semi-classical model that is either non-linear, non-energy conserving or violates superluminal signalling. As we have clarified here, this conclusion carries over to generic models of gravitational radiation interacting with matter, and therefore, to the gravito-phononic effect \cite{tobar2023detecting}. In this way, any consistent model of energy exchange of single quanta between gravitational radiation and quantum matter, in analogy with the photo-electric effect for photons, must involve the quantisation of gravitational radiation.

\section{Conclusion}
We revisited the photo-electric effect using semi-classical approaches and full quantum electrodynamics, and extended these insights to the gravito-phononic effect of  Ref.~\cite{tobar2023detecting}. By exploring energy transfer between coupled harmonic oscillators, we outlined that key signatures traditionally associated with the photo-electric effect-such as threshold frequencies, intensity-independent transitions, and discrete quanta absorption-can be replicated with coupled harmonic oscillators. This reveals that coupled oscillator systems can serve as effective analogues for studying quantized energy exchange between bosonic modes. This cements gravitational wave detectors as systems for implementing a gravitational version of the photo-electric effect. We further highlighted the role of energy conservation in these systems, showing that while semi-classical and classical field models can mimic quantum signatures, they face challenges in conserving energy at the level of single transitions. Such pathologies can be minimized with non-linearities, but this comes at the expense of substantial modifications to quantum mechanics. This underscores the necessity of quantising the linearised gravitational field, for a consistent explanation of the experimental implementation of the gravito-phononic effect proposals in Refs.~\cite{tobar2023detecting,tobar2024detectingkhzgravitonsneutron}. Furthermore, as the parametric conversion of gravitons to photons in electromagnetic detectors, which are subsequently read-out by a photon-counter detector can similiarly be described by a beampslitter interaction, extensions to electromagnetic detectors of Refs.~\cite{PhysRevLett.74.634,AlexanderDolgov,ADDAZI2024138574,ADDAZI2025101844} follows. Therefore, implementing the proposals of Refs.~\cite{tobar2023detecting,tobar2024detectingkhzgravitonsneutron} - which involve graviton detectors targeting known sources of gravitational radiation, with established gravitational wave detection methods, will yield the most substantial experimental evidence for the quantisation of gravity to date.

\onecolumngrid
\appendix
  \section{Beamsplitter dynamics}\label{beamappendix}
Here we derive the expressions given in Section \ref{beamsplitterdynamics} for the dynamics of the beamsplitter interaction of two interacting bosonic modes. The Hamiltonian for a bosonic mode of frequency $\nu$ (which we model as the gravitational field), and a bosonic mode of frequency $\omega$ (which we model as the detector) is
\begin{equation}
    \hat{H}=\hbar \nu \hat{a}^{\dagger} \hat{a}+\hbar \omega \hat{b}^{\dagger} \hat{b}+\hbar g\left(\hat{a} \hat{b}^{\dagger}+\hat{b} \hat{a}^{\dagger}\right),
\label{Xeqn24-A.1}
\end{equation}
with coupling strength $g$. Transforming into the rotating frame of the free Hamiltonian $\hat{H}_0$, we obtain:
\begin{equation}
    \hat{H}_0 = \hbar \nu \hat{a}^{\dagger} \hat{a}+\hbar \omega \hat{b}^{\dagger} \hat{b},
\label{Xeqn25-A.2}
\end{equation}
where the interaction Hamiltonian is 
\begin{equation}
    \hat{H}_\mathrm{int} = \hbar g\left(\hat{a} \hat{b}^{\dagger}e^{-i\left(\omega-\nu\right) t} + e^{-i\left(\nu-\omega\right) t}\hat{b} \hat{a}^{\dagger}\right).
\label{Xeqn26-A.3}
\end{equation}
We now consider a coherent state of high amplitude of the gravitational field interacting with a ground state cooled resonant mass detector of frequency $\omega$, and we have that $|\psi(t)\rangle_I=e^{i \hat{H}_0 t}|\psi(t)\rangle_{S} =U_I(t)\left|\psi_I(0)\right\rangle$, where 
\begin{equation}
\begin{split}
    U_I(t) &= \mathcal{T}\left\{e^{-i \int_0^t d t^{\prime} \hat{H}_{\mathrm{int}}\left(t^{\prime}\right)}\right\} \\
    &= \hat{\mathbb{I}}-i \int_0^t d t^{\prime} \hat{H}_\mathrm{int}\left(t^{\prime}\right) + \mathrm{... },
\end{split}
\label{Xeqn27-A.4}
\end{equation}
We truncate at first order in the expansion and examine the probability to transition to the to the excited state, which is evaluated as 
\begin{equation}
\begin{split}
    P(\ket{n = 1}) &= g^2|\alpha|^2\int_0^t d t^{\prime}\int_0^t d t^{\prime\prime} e^{-i(\omega-\nu) (t' -t'')} \\
    &= 4g^2|\alpha|^2\frac{\sin ^2\left(0.5\left(\nu-\omega\right) t\right)}{\left(\nu-\omega\right)^2}.
\end{split}
\label{Xeqn28-A.5}
\end{equation}

\section{Interaction Hamiltonian in the TT gauge}
Here we derive the interaction Hamiltonian of gravitational radiation with a weber bar - the most commonly used harmonic oscillator model for a detector of gravitational radiation, in the TT gauge. When the toy model discussed in the main text, is transformed into the gravitational radiation model, this is the explicit form of the gravitational radiation interacting with quantum matter that is used - gravitational radiation interacting with the collective mode of a mass (as a quantum harmonic oscillator), in the TT-gauge. The interaction Hamiltonian between linearised gravity and matter is
\begin{equation}
    H_{\mathrm{int}}=-\frac{1}{2} h_{\mu \nu} T^{\mu \nu} ,
\label{Xeqn47-C.1}
\end{equation}
where $T^{\mu \nu}$ is the stress energy tensor and $h_{\mu \nu}$ is the metric perturbation to the Minkowski metric in the linearised limit. A common choice of gauge considered in the context of graviton absoprtion is the \textit{local-inertial frame}, in which the generalised velocities are negligible, reducing the interaction Hamiltonian to be
\begin{equation}
    H \approx-\frac{1}{2} T^{00} h_{00},   
\label{Xeqn48-C.2}
\end{equation}
which can be further simplified with $T^{00} = m$ for a mass element m. Now $h_{00}$ can be expressed in terms of the TT gauge plane waves. Using Fermi-normal co-ordinates 
\begin{equation}
    h_{00} = -R_{0 j 0 k} x^j x^k.
\label{Xeqn49-C.3}
\end{equation}
Using the definition of the Riemann tensor, the interaction Hamiltonian in the local inertial frame can be expressed as
\begin{equation}
    H = \frac{m}{4} x^j x^k  \ddot{h}^{\mathrm{TT}}_{jk} .
\label{Xeqn50-C.4}
\end{equation}
In this case, the force on the mass element $m$, can be expressed as:
\begin{equation}
    F_i = \frac{\partial H}{\partial x_i} = \frac{m}{2}\ddot{h}^{\mathrm{TT}}_{il}x^l.
\label{Xeqn51-C.5}
\end{equation}
This force is what is subsequently used on the individual mass elements that make up the bar resonator to derive the effective interaction. This has been expressed in terms of the TT gauge plane waves, using Fermi-normal co-ordinates. Following \citet{MaggioreMichele2007GwV1}, the strain field of a bar antenna $\boldsymbol{u}(x, y, z, t)$ follows the equation of motion
\begin{equation}
    \rho \ddot{u}_\alpha - \lambda \frac{\partial^2}{\partial x_\alpha \partial x_\beta} u_\beta=\frac{1}{2} \rho  \ddot{h}_{\alpha \beta} x_\beta,
\label{Xeqn52-C.6}
\end{equation}
where $\rho$ is the density of the material, while $\lambda$ is the elasticity coefficients of the material. The term on the RHS describes the previously derived force on the individual mass elements in the \textit{local-inertial frame}. Expanding the strain field into a series of eigenmodes, 
\begin{equation}
    \boldsymbol{u}_x(x, y, z, t)=\sum_n b_n(t) \boldsymbol{w}_n(x, y, z).
\label{Xeqn53-C.7}
\end{equation}
We now restrict to the simple case of 1-D acoustic oscillations and consider sinusoidal strain-fields:
\begin{equation}
    u_x(t, x)=\sum_{n}  b_{n }(t) \sin \left[\frac{\pi x}{L}(2 n+1)\right] ,
\label{Xeqn54-C.8}
\end{equation}
before substitution, we set $\mu$ = 0, and identify the speed of sound in the material as $v_s = \sqrt{\frac{\lambda}{\rho}}$, which for 1-D oscillations (and only considering the coupling to the $xx$ component of the gravitational wave for simplicity), we obtain:
\begin{equation}
\ddot{u}_{x} - v_s^2  \frac{\partial^2}{\partial x^2} u_{x} = \frac{1}{2}   \ddot{h}_{xx} x.
\label{Xeqn55-C.9}
\end{equation}
We now substitute in the mode expansion, and then using the orthogonality relation 
\begin{equation}
    \int\limits_{-L / 2}^{L / 2} \mathrm{d}x \; \sin{\left[\frac{\pi x}{L}(2 n+1)\right]} \sin{\left[\frac{\pi x}{L}(2 m+1)\right]} = \frac{L}{2} \delta_{n,m},
\label{Xeqn56-C.10}
\end{equation}
the equation of motion simplifies to the following ODE for the normal modes $b_n(t)$:
\begin{equation} \label{eom111}
\ddot{b}_{n}(t)+b_{n }(t)\left[v_s^2\left(\frac{\pi(2 n+1)}{L}\right)^2\right]=\frac{(-1)^n}{(2 n+1)^2}\left(\frac{2 L}{\pi^2}\right) \ddot{h}_{x x},
\end{equation}
where we can identify 
\begin{equation}
    \omega_n = v_s^2\left(\frac{\pi(2 n+1)}{L}\right)^2,
\label{Xeqn58-C.12}
\end{equation}
as the frequency of the normal mode. This is simply the equation of a simple harmonic oscillator driven by the force 
\begin{equation}
    F = M\frac{(-1)^n}{(2 n+1)^2}\left(\frac{2 L}{\pi^2}\right) \ddot{h}_{x x},
\label{Xeqn59-C.13}
\end{equation}
 The above force can now be converted to a Hamiltonian through $H = -\partial F /\partial x$, which results in the following interaction Hamiltonian (as first considered by \citet{MaggioreMichele2007GwV1}):
\begin{equation} \label{hintqg}
    H_{\mathrm{int}} = \frac{(-1)^n}{(2 n+1)^2}\left(\frac{2ML}{\pi^2}\right) \ddot{h}_{x x}{b}_{n}.
\end{equation}
Giving the interaction Hamiltonian we use in the toy model in the main text, replacing the phononic mode with the generic oscillator mode. For a gravitational wave propagating along the $z$ axis, we use the gravitational field in terms of Fourier amplitudes as
\begin{equation} \label{hmunu}
    h(t) =  \sum_k \left(  h_\mathbf{k} e^{i (k z - \Omega t)} + h_\mathbf{k}^* e^{-i (k z + \Omega t)} \right).
\end{equation}
which we use in the main text. We now proceed to quantising the linearised gravitational field used here, in terms of plane waves, identifying the fourier amplitude $h_{ \mathbf{k}} = \frac{1}{c} \sqrt{\frac{8 \pi G \hbar}{V \nu_{\mathbf{k}}}}\hat{a}_{\mathbf{k}}(t)$. In the main text, we use this quantisation of the quantisation of the linearised gravitational field.

\bibliography{refs}

\end{document}